\documentclass[a4paper,12pt]{article}
\usepackage[utf8]{inputenc}
\usepackage{amsmath}
\usepackage{amssymb}
\usepackage{mathrsfs}
\usepackage{amscd}
\usepackage{chemarrow}
\usepackage{times}
\usepackage{graphicx}

 \newcounter{thm}
 \newcounter{ex}
 \newcounter{re}
\def\rd{{\rm d}}

\newcommand{\x}{\ensuremath{\mathbf{x}}}

 \newcommand{\R}{\ensuremath{\mathbb{R}}}
\newcommand{\Z}{\ensuremath{\mathbb{Z}}}

\newcommand{\absval}[1]{\ensuremath{\left|#1\right|}}

\newcommand{\pderiv}[3][]{\ensuremath{\frac{\partial^{#1}#2}{\partial{#3}^{#1}}}}

\renewcommand{\d}{\ensuremath{\,\textrm{d}}}
\DeclareMathOperator{\argmax}{argmax}

\begin{document}

\title{Nonlinear Stochastic Dynamics of Complex Systems, II:
Potential of Entropic Force in
Markov Systems with Nonequilibrium Steady State,
Generalized Gibbs Function and Criticality
}

\author{Lowell F. Thompson and Hong Qian 
\\[10pt]
Department of Applied Mathematics\\
University of Washington\\
Seattle, WA 98195-3925, U.S.A
}

\maketitle




\abstract{In this paper we revisit the notion of the ``minus logarithm of stationary probability'' as a generalized potential in nonequilibrium systems and attempt to illustrate its central role in an axiomatic approach to stochastic nonequilibrium thermodynamics of complex systems.  It is demonstrated that this quantity arises naturally through both monotonicity results of Markov processes and as the rate function when a stochastic process approaches a detrministic limit.  We then undertake a more detailed mathematical analysis of the consequences of this quantity, culminating in a necessary and sufficient condition for the criticality of stochastic systems.  This condition is then discussed in the context of recent results about criticality in biological systems.   }










\section{Introduction}

	This is part II of a series on stochastic nonlinear dynamics of 
complex systems. Part I \cite{qian-part-1} presents a 
chemical reaction kinetic perspective on complex 
systems in terms of a mesoscopic stochastic nonlinear kinetic 
approach, e.g., Delbr\"{u}ck-Gillespie processes, as well as a 
stochastic nonequilibrium thermodynamics (stoc-NET) in phase space. 
One particularly important feature of the theory in \cite{qian-part-1}
is that it takes the abstract mathematical concepts seriously --
that is, it follows what the mathematics tells us.  For example, it was shown that 
the widely employed {\em local equilibrium assumption} in the 
traditional macroscopic theory of NET can be eliminated when one recognizes the 
fine distinction between the set of random events, the $\mathscr{S}$ 
in a probability space $(\mathscr{S},\mathcal{F},P)$ and a random 
variable that is defined as an observable on the top of the measurable 
space, $\x: \mathscr{S}\rightarrow \mathbb{R}$.  The local
equilibrium assumption is needed only when one applies the 
phase space stoc-NET to physically measurable transport 
processes \cite{qkkb-16}.

	The same chemical kinetic approach can be applied to other
biological systems.  Biological organisms are complex systems with a 
large number of
heterogeneous constituents, which can be thought of as ``individuals''.
To be able to develop a scientific theory for such a complex
system with any predictive power, one must use a probabilistic treatment that 
classifies the individuals into ``statistically identical groups''.  Thermodynamics and statistical mechanics provide 
a powerful conceptual framework, as well as a set of tools with which 
one can comprehend and analyze these systems.  The
fully developed statistical thermodynamic theory taught in college 
physics classes is mainly a theory of equilibrium systems.  The
application of its fundamental ideas, however, is not limited to just equilibrium 
systems or molecular processes.  Stoc-NET \cite{qkkb-16,vdb-e,zqq,seifert,jarzynski}, along with the information theoretical approach \cite{jaynes-book,bialek-1,bialek-2,bialek-3}, is a further development in this area.

 	One of the key elements of the theory presented in \cite{qian-part-1} 
was the nonequilibrium steady state (NESS) potential, or ``energy'', 
defined as the minus logarithm of the stationary probability distribution of 
a kinetic model.  In the past, this quantity has appeared repeatedly in 
the literature \cite{haken-graham,kubo,nicolis-lefevere,yin-ao,feng-wang-11}, but
most of the studies focus on its computation.  In this 
paper, we attempt to illustrate its central role as a novel 
``law of force'', a necessary theoretical element in the stoc-NET 
of complex systems.

	The paper is organized as follows:  Section \ref{sec:2} serves as a brief historical review of the use of the negative logarithm of a stationary probability distribution as an energy potential.  In Sec. \ref{sec:2.1} we first
look at the history of using minus-log-probability to equilibrium chemical thermodynamics and 
briefly review J. G. Kirkwood's fundamental idea of the potential of mean
force and the notion of entropic force.  In Sec. \ref{sec:2.2},
two recent results identifying the minus-log-probability as
``energy'' are described: a self-contained and consistent mesoscopic 
stoc-NET \cite{ge-qian-10}, and a precise agreement between its 
macroscopic limit and Gibbs' theory \cite{ge-qian-16,ge-qian-16b}.  These two results provide strong evidence for the validity of such an identification.  
In Sec. \ref{sec:2.3}, we discuss the legitimacy and centrality of 
stationary distribution in the ``entropy inequality'' for a 
Markov process from a mathematical standpoint. 

In Sec. \ref{sec:3}, a definition of the ``corresponding deterministic dynamics'' of a stochastic process is proposed using power-scaling of probability densities.  In Sec. \ref{sec31} it is shown that the rate of convergence to this corresponding deterministic process coincides with the minus-log-probability definition of energy.  With the justifications
given in Sec. \ref{sec:2} and \ref{sec:3}, a more detailed analysis of such a probability 
distribution is carried out in Sec. \ref{sec:criticality}.  In Sec. \ref{sec41} terms analagous to Boltzmann's and Gibbs' entropy are defined, along with their corresponding microcanonical partition functions.  The relative merits of these definitions are discussed.  In Sec. \ref{crit-sec}, it is shown that the system has a crtical temperature if and only if the Gibbs' entropy of the system is asymptotic to the energy.  In Sec. \ref{sec43} several example distributions are discussed in order to emphasize some subtleties in the definition of states.  Finally, in Sec. \ref{sec:5} the ideas from previous sections are related to some recent results on
biological systems.  

\section{A novel law of force: Potential of entropic force}
\label{sec:2}

In Boltzmann's statistical mechanics, phenomenological 
thermodynamics is given a Newtonian mechanical basis.  
Based on the already well developed concepts of mechanical energy and its conservation, Boltzmann derived the relation\footnote{Boltzmann's mathematical
derivation matched the modern {\em maximum entropy
principle} with the constraint of given mean value
for energy, which yields an exponential law for the enegy 
distibution.  Note the mathematical statements of
energy conservation $\sum_{k=1}^N E_k = C$ and
fixed mean energy $\tfrac{1}{N}\sum_{k=1}^N E_k
= \overline{c}$ are equivalent when $N$ is given.} 

\begin{equation}
                    p^{eq}(x)\propto e^{-U(x)/k_BT},
\label{b-law}
\end{equation}
where $U(x)$ is the mechanical energy of a microstate\footnote{However, a thermodynamic state is a state of recurrent
motion; defined by an entire level set $\mathcal{A} = \{x \:\vert\: U(x)=E\}$.  
Thus, Boltzmann also introduced his celebrated entropy $S_{B}(E)=k_B\ln\Omega(E)$ 
where $S_{B}$ is the entropy and $\Omega(E)$ is the number of 
microstates consistent with a given energy $E$.  That is, $\Omega(E)$ is the cardinality of $\mathcal{A}$.  In 
terms of $E$, then $p^{eq}(E) \propto \Omega(E)e^{-E/k_BT}
=e^{-[E-TS(E)]/k_BT}$.} $x$ 
and $p^{eq}(x)$ is the probability of state $x$ when the system is in 
{\em thermal equilibrium} -- a concept which had also already
been well established in thermodynamics via the notion of
{\em quasi-stationary processes}.   In a thermodynamic equilibrium, 
there is no net transport of any kind.\footnote{In the thermodynamics before Gibbs, macroscopic transport processes were driven by either
a temperature or a pressure gradient in the three-dimensional 
physical space.  In Gibbs' macroscopic chemical thermodynamics,
a chemical equilibrium has no net flux in the abstract 
stoichiometric network.  In the current mesoscopic, stochastic
thermodynamics, an equilibrium has no net probability transport
in an appropriate state space.  The notion of detailed balance
independently arose in physics \cite{maxwell-67, boltzmann-72}, chemistry \cite{wegscheider, gnlewis-25} and in 
probability theory \cite{kolmogorov-36}. }

	Inspired by Boltzmann's law (\ref{b-law}), generalizations of
the concept of equilibrium thermodynamic potentials have been
proposed in many studies.  These generalizations go by a variety 
of names: generalized thermodynamic potential, kinetic potential, 
nonequilibrium potential, pseudo-potential, emergent 
landscape, etc. \cite{haken-graham,kubo,nicolis-lefevere,yin-ao,feng-wang-11,qian-ge-mcb}.  One of the common features of all these names is that the ``potential function'' is defined by 
applying Eq. \ref{b-law} in reverse.  One \emph{defines} a
potential 

\begin{equation} \label{eq:Hdef}
H(x) = -\ln p^{eq}(x)
\end{equation}
based on the stationary probability, which can be obained
in many statistical models and whose existence can be
mathematically proven for a large class of systems.  
Most importantly, many systems with stationary probability have
non-zero transport flux(es)!

	In fact, this tradition of taking \eqref{eq:Hdef} as a 
legitimate potential function started in equilibrium
statistical chemical thermodynamics.  Note that according to 
Eq. \ref{b-law}, the term $-k_BT\ln p^{eq}(x)$ is simply the
total mechanical energy of state $x$, which is known
{\em a priori}.  Therefore, there is no reason to define \eqref{eq:Hdef}
in studies of a pure mechanical system.  However, in
statistical {\em chemical} thermodynamics, one usually does
not have a full Hamiltonian function for a 
complex molecule in hand.  It is at this juncture that the
notion of a {\em potential of mean force} \cite{kirkwood-35}
enters the theory.

\subsection{Equilibrium potential of mean force}
\label{sec:2.1}

	Physical chemists deal with complex molecules and force fields.
Even though in molecular dynamics (MD) a 
molecule has a classical mechanical
representation in terms of atoms as point masses, 
the precise potential energy is not known. 
The force fields in MD have therefore been under intense 
development over the past 50 years \cite{levitt}.   With such complexities, 
is it even possible to do statistical mechanics?

	Let us first note a very important mathematical 
equality in connection to Eq. \ref{b-law}.  We consider
a function $U(x)$ with $x=(x_1,x_2)$ where $x\in\mathscr{S}=\mathscr{S}_1\oplus \mathscr{S}_2$,
$x_1\in\mathscr{S}_1$ and $x_2\in\mathscr{S}_2$.  Then,
\begin{eqnarray}
		Z(T) &=& \int_{\mathscr{S}} e^{-U(x)/k_BT}\rd x
\nonumber\\
	 &=& \int_{\mathscr{S}_1}\int_{\mathscr{S}_2} e^{-U(x_1,x_2)/k_BT}\rd x_1\rd x_2
\nonumber\\
	 &=& \int_{\mathscr{S}_1} e^{-\varphi(x_1)/k_BT}
              \rd x_1, 
\label{b-law-like}
\\
	\varphi(x_1) &=& -k_BT\ln\int_{\mathscr{S}_2}
           e^{-U(x_1,x_2)/k_BT}\rd x_2.
\label{pomf}
\end{eqnarray}
We see that if one considers $\varphi(x_1)$ as a
``potential function'' for the system in (coarse-grained) state
$x_1$, then one can obtain the same $Z(T)$ using 
Eq. \ref{b-law-like}, which is in the exact same form as in
(\ref{b-law}).  More importantly, one sees that $\varphi(x_1)$
is the free energy with fluctuating $x_2$ and fixed $x_1$.

	After reading the calculations above, one is naturally led to the question, ``what does this potential energy function $\varphi(x_1)$ 
defined in (\ref{pomf}) represent?'' J. G. Kirkwood answered this 
question in a very satisfying manner \cite{kirkwood-35}:  
it is the potential function of a ``mean force'', in equilibrium, 
acting on the system which is fixed at $x_1$.
\begin{eqnarray}
		-\frac{\rd\varphi(x_1)}{\rd x_1} &=&
       -\frac{\displaystyle \int_{\mathscr{S}_2}
            \left(\frac{\partial U(x_1,x_2)}{\partial x_1}\right)_{x_2}
           e^{-U(x_1,x_2)/k_BT}\rd x_2 }{\displaystyle \int_{\mathscr{S}_2}
           e^{-U(x_1,x_2)/k_BT}\rd x_2 }
\nonumber\\[6pt]
	&=&  -\int_{\mathscr{S}_2}
            \left(\frac{\partial U(x_1,x_2)}{\partial x_1}\right)_{x_2}
                  p^{eq}\big(x_2 \big| x_1\big)\rd x_2,
\label{the-mf}
\end{eqnarray}
in which $p^{eq}(x_2 | x_1)$ is the conditional
equilibrium probability distribution for $x_2$ given $x_1$,
and the partial derivative 
$-(\partial U(x_1,x_2)/\partial x_1)_{x_2}$
is precisely the mechanical force in the $x_1$ direction,
with the given $x_2$.  Averaging over the fluctuating $x_2$
with distribution $p^{eq}(x_2|x_1)$, Eq. \ref{the-mf} is 
the mean force on $x_1$.

	In other words, Eq. \ref{pomf} states that 
the negative logarithm of the marginal probability distribution
for $x_1$ is simply the potential of mean force if one chooses the free energy
of the entire system, $F(T)=-k_BT\ln Z(T)$, as the zero energy reference point.  

\begin{equation}
   -k_BT\ln \int_{\mathscr{S}_2} p^{eq}(x_1,x_2)\rd x_2 
      = \varphi(x_1) - F(T).
\end{equation}
One of the most important facts, as is clear from 
(\ref{pomf}), is that the 
potential of mean force $\varphi(x_1)$ is itself a
function of temperature.  In physical chemistry,
one usually builds a statistical mechanical model using such a 
potential of mean force rather than using a mechanical energy function.  That is, one uses a free energy function with
certain degrees of freedom fixed, and averaged over all
the others.

	Since $\varphi(x_1)$ is temperature dependent,
it has its own energy part and entropy part:

\begin{equation}
   \varphi(x_1;T) = \underbrace{\frac{\partial (\varphi/T)}{\partial (1/T)} }_{\text{energy}}
          - T \underbrace{\left(-\frac{\partial \varphi}{\partial T}  \right) }_{\text{entropy}}
\end{equation}
A potential of mean force can be purely entropic.
One of the best known examples is rubber elasticity, which 
arises from a Gaussian polymer chain \cite{hill-sm-book}.
If the temperature is suddenly droped to zero, the force
(and its associated energy) disappears instantly. 

Observing this significant conceptual distance between chemical thermodynamics 
and its mechanical origin, and the essential statistical nature of Gibbs' 
energy based on minus-log probability in all modeling practices, it is not 
surprising that some researchers who mainly work with biochemical 
thermodynamics strongly feel that one could reformulate statistical thermodynamics (at least in connection to energy) in terms of a ``measure of information'' and abandon the very term ``entropy'',
along with its root in mechanics \cite{ben-naim}.

\subsection{Nonequilibrium steady state potential}
\label{sec:2.2}

	For stochastic models of equilibrium systems, therefore, \eqref{eq:Hdef}
yields a meaningful free energy function, in $k_BT$ units.
It embodies an exact coarse-graining procedure.
For stochastic models of nonequilibrium steady
state with non-zero transport flux, we now have sufficient
evidence to suggest that 

\begin{equation} \label{eq:H}
H(x) = -\ln p^{ss}(x)
\end{equation}
(where $p^{ss}$ is a stationary distribution, but may or may not be an equilibrium distribution) is also a meaningful energy function.  We 
start with some conceptual discussions.

	First, outside classical mechanics, the question ``what is a force and
how do we quantify it'' is highly non-trival and vague.  
Onsager, however, introduced the notion of a 
{\em thermodynamic force} in his theory of irreversible 
processes \cite{onsager-31}.  Intuitively, a force is the 
cause of an action.  In Newtonian mechanics, a force is the 
cause of a change in the vector $\frac{\rd}{\rd t}\vec{x}$.
But in an ``overdamped world'', which emcompasses most of 
chemistry, biology, and society, a force is actually needed 
to cause a meaningful movement (i.e., a transport).

In terms of the mathematical theory of stochastic 
dynamics, there is a universal 
conception for movement, or ``dynamics'':
{\bf\em Given the option to move to one of many states, a system is most likely to move to the state with the highest stationary probability.}
One should immediately note
that this statement is highly problematic from a 
rigrous mathematical standpoint.  Nevertheless, 
at least in one class of systems, the above notion
is attainable:  the class of systems whose
dynamics have an invariant measure that is 
ergodic.

	When discussing statistical mechanics, Montroll and
Green have stated that \cite{montroll-green} 
``The aim of statistical mechanics is to develop a formalism from which
one can deduce the macroscopic behavior of physical systems composed of a large number of molecules from a specification of the component molecular species, the laws of force which govern intermolecular interactions, and the nature of their surroundings.''
With the rise of equilibrium chemical thermodynamics, it is clear that 
the ``laws of force'' themselves can be discovered from the
equilibrium distribution.  In fact, most such laws of force in 
biophysical modeling are statistical in nature and can be seen as entropic forces.

	Indeed, ``[t]o date no one has succeeded in deriving the laws of nonequilibrium phenomena from the [Newtonian] equations of motion merely by allowing the number of particles involved to become infinite. However, considerable success has been achieved by introducing various statistical hypotheses.'' \cite{montroll-green}
Recent studies have shown that if one identifies $H(x)$
as a ``generalized Helmhotz or Gibbs energy function'', 
a complete and consistent
mesoscopic thermodynamics can be formulated that includes
nonequilibrium steady states \cite{ge-qian-10,qkkb-16}.  Furthermore,
if one passes the system from mesoscopic to macroscopic by
allowing the number of particles involved and the 
system's volume to become infinite, two macroscopic 
thermodynamic laws can be derived \cite{ge-qian-16}.  If 
the mesoscopic system is a general chemical reaction network
with detailed balance, the macroscopic emergent potential was
shown mathematically to be Gibbs' function $G(x)$,
where $x_i$ are the concentrations of chemical species, with
$\partial G/\partial x_i$ being the chemical potential
for the $i^{th}$ species.  The same theory also proves
the existence of, and provides an equation for computing, 
a generalized Gibbs function for an open chemical reaction 
network under a chemostat, which approaches to 
a nonequilibrium steady state.

\subsection{Stationary distribution and entropy inequalities of
Markov processes}
\label{sec:2.3}

	Unless stated otherwise, we will exclusively deal with a denumerable 
state space $\mathscr{S}$ (either finite or infinite) for the remainder of the 
paper.

{\bf\em A stronger monotonicity result.}
The strongest version of a monotoic entropy result that we 
are aware of is \cite{lindblad-73,voigt-81}

\begin{equation}
  \frac{\rd}{\rd t}
D\big[\{p_x(t)\}\|\{q_x(t)\}\big] \equiv \frac{\rd}{\rd t}
          \sum_{x\in\mathscr{S}}
         p_x(t) \ln\left(\frac{p_x(t)}{q_x(t)}\right) \le 0,
\label{voigt}
\end{equation}
in which $p_x(t)$ and $q_x(t)$ are two solutions to the 
Kolmogorov forward equation with different initial 
distributions.  Eq. \ref{voigt} immediately yields a variety of 
related inequalities:

(i) When $q_x(t)\equiv \pi_x$ $\forall t$,
where $\{\pi_x\}$ is a stationary distribution of the Markov
process, then (\ref{voigt}) is the widely known 
``free energy theorem'' \cite{cover_book,qian-pre-01}.

(ii) When $q_x(t)\equiv \pi_x$ $\forall t$, and $p_i(0)=\delta_{i\ell}$,
one has

\begin{subequations}
\begin{equation}
    \frac{\rd}{\rd t}
          \sum_{j\in\mathscr{S}}
         p_{\ell j}(t) \ln\left(\frac{p_{\ell j}(t)}{\pi_j}\right) \le 0
 \ \forall \ell,
\end{equation}
therefore,

\begin{equation}
  \frac{\rd}{\rd t} I\big[\x_t\big\|\x_0\big]  = \frac{\rd}{\rd t}
          \sum_{\ell,j\in\mathscr{S}}
         \pi_{\ell}p_{\ell j}(t) \ln\left(\frac{p_{\ell j}(t)}{\pi_{\ell}\pi_j}\right) \le 0;
\end{equation}
where $I[\x_t\big\|\x_0]$ is the mutual information between
$\x_0$ and $\x_t$ of a stationary Markov process.  Similarly,

\begin{equation}
 \frac{\rd}{\rd t}\left(
          -\sum_{\ell,j\in\mathscr{S}}
         \pi_{\ell}p_{\ell j}(t) \ln p_{\ell j}(t)\right) \ge 0.
\end{equation}
This result was in \cite{yubin-2008}.  The term inside $(\cdots)$ 
is the conditional Shannon entropy $H[\x_t|\x_0]$ for the
stationary $\x_t$.  It is also the Kolmogorov-Sinai (KS) entropy of 
every $t$ steps of the stationary $\x_t$:

\[
          \lim_{n\rightarrow\infty}
              \frac{1}{n}H\big[\x_0,\x_t,\x_{2t},\cdots \x_{nt} \big].
\]
The result is more easily understood when interpreted this way: KS entropy 
quantifies the randomness in a ``map''.  The randomness
does not decrease with map composition.
\end{subequations}

(iii) When $p_x(t)\equiv\pi_x$ (and when we then rename $q_x(t)$ as $p_x(t)$),
we have

\begin{equation}
  \frac{\rd}{\rd t}
          \sum_{x\in\mathscr{S}}
         \pi_x \ln \frac{\pi_x}{p_x(t)} \le 0.
\label{eq9}
\end{equation}
To explain this result more intuitively, we note that the sum in \eqref{eq9} can be interpreted as the information lost when predicting $\pi_x$ from $p_x(t)$.  Roughly speaking, if $t_1 < t_2$, then it takes more information to predict the distant future ($\pi_x$) from time $t_1$ than it does from time $t_2$ because the prediction from $p_x(t_1)$ has to account for the random events that can happen within the time interval $[t_1,t_2]$.

	{\bf\em Filtration and entropy monotonicity.}
Even though the original Shannon
entropy used an implicit uniform prior, the necessity for an
explicit prior has been widely discussed in information theory\footnote{The entropy with respect to an explicit prior is more accurately called the relative entropy or cross-entropy, and its expression is analagous to the free energy in statistical mechanics.} 
\cite{hobson,shore-johnson}.   More importantly, for a
continuous random variable, the logarithm of a probability density is 
simply ill-defined mathematically.    All the various 
monotoic ``entropy'' results in the previous section 
provide the legitimacy of using $\{\pi_x\}$ as the reference 
measure for a Markov process.  We would like to argue
that this is in fact necessary.

	We consider a Markov process in a more general setting 
in this section.  Let the triple $(\mathscr{S}, \mathcal{F}, P)$ 
be a probability space; let $(\mathcal{I}, \le)$ be a totally 
ordered index set; and let $(S, \Sigma)$ be a measurable space.
If
$X:\mathcal{I}\times\mathscr{S} \rightarrow S$ 
is a stochastic process, then its natural filtration of $\mathcal{F}$ 
with respect to $X$ is a 
sequence $\big\{\mathcal{F}_i^{(X)} \:\vert\: i\in\mathcal{I}\big\}$ 
such that

\begin{equation}
    \mathcal{F}_i^{(X)} = \sigma\Big\{X^{-1}_j(A)\ \:\Big|\: \  
                 j\in\mathcal{I}, j\le i, A\in\Sigma\Big\}.
\label{f-eq0001}
\end{equation}
That is, $\mathcal{F}_i^{(X)}$ is the smallest $\sigma$-algebra on $\mathscr{S}$
that contains all pre-images of $\Sigma$-measurable subsets
of $S$ for times $j$ up to $i$.
The definition given in (\ref{f-eq0001}) yields a
monotonic relation

\begin{equation}
     \mathcal{F}^{(X)}_j \subseteq \mathcal{F}^{(X)}_i \  
          \textrm{ if } \   i, j\in\mathcal{I}, \  j\le i.
\label{filtration-002}
\end{equation} 
Such a property is called {\em non-anticipating};
in other words,  ``when including the future, the dynamics 
are at least as random as up to now.''  

The monotonicity in
Eq. \ref{filtration-002} can be expressed in terms of 
Shannon's information entropy as

\begin{equation}
    H[X_0,X_1,\cdots,X_i]\le H[X_0,X_1,\cdots,X_i,X_{i+1}].
\label{filtration-003}
\end{equation}
This inequality is true because 
$H[X_0,\cdots,X_{i+1}]-H[X_0,\cdots,X_i]$
is the conditional Shannon entropy $H[X_{i+1}|X_1,\cdots,X_i]$,
which is never negative.

	Notice that Eqs. \ref{filtration-002} and \ref{filtration-003}  
are concerned with the \emph{sequences} of $\big\{X_j \:\vert\: j\le i\big\}$, but the ``entropy 
monotonicity'' results in statistical physics deal with \emph{individual} $X_i$ 
and $X_{i+1}$, and entropy has deterministic values that are
different for different times.  The relationship among
$X_i$, $X_{i+1}$, and the filtration is shown as 

\begin{equation}
\CD
   (\mathscr{S},\mathcal{F}) @ > X_i >>  (S,\Sigma) @ > X_i^{-1} >>   \Big(\mathscr{S},F_i^{(X)}\Big)  \\
    @|      @|      @V  \text{time stepping} VV  \\
  (\mathscr{S},\mathcal{F}) @ > X_{i+1} >>  (S, \Sigma)  @> X_{i+1}^{-1} >>  \Big(\mathscr{S},F_{i+1}^{(X)}\Big)
\endCD
\end{equation}

	We now consider the information
lost from $X_i$ to $X_{i+1}$ when the event $\omega$ occurs,
$\ln P_{X_{i+1}}(\omega)-\ln P_{X_i}(\omega)$.  Then its 
expected value with respect to the stationary, invariant measure $\mu_{\pi}(\omega)$ is given by 

\begin{eqnarray}
     \mathbb{E}\left[\ln P_{X_{i+1}} - \ln P_{X_i}\right] &=& \int_{\Omega}
      \ln\left(\frac{\rd P_{X_{i+1}}}{\rd P_{X_i}}
              (\omega)\right) \rd\mu_{\pi}(\omega)
\nonumber\\
    &=& \int_{\Omega}\ln\left(\frac{\rd P_{X_{i+1}}}{\rd\mu_{\pi}}(\omega)\right)  \rd\mu_{\pi}(\omega)
       - \int_{\Omega} \ln\left(\frac{\rd P_{X_i}}{\rd\mu_{\pi}}(\omega)\right)  \rd\mu_{\pi}(\omega).
\label{filtration-004}
\end{eqnarray}
If both $X_i$ and $X_{i+1}$ are real valued (i.e.,
$S=\mathbb{R}$) with density functions $f_{X_i}(x)$ and
$f_{X_{i+1}}(x)$ respectively,  then (\ref{filtration-004})
becomes

\begin{equation}
    \mathbb{E}\left[\ln P_{X_{i+1}} - \ln P_{X_i}\right] = \int_{\mathbb{R}}\ln\left(\frac{f_{X_{i+1}}(x)}{\pi(x)}\right)\pi(x)\rd x
       - \int_{\mathbb{R}}
   \ln\left(\frac{f_{X_i}(x)}{\pi(x)}\right)  \pi(x)\rd x,
\label{filtration-005}
\end{equation}
where $\pi(x)=\rd\mu_{\pi}/\rd x$ is the density of the 
stationary measure.  
We know that Eq. \ref{filtration-005} is never negative; therefore
the mean information lost

\begin{equation}
              \int_{\Omega}
      \ln\left(\frac{\rd P_{X_{i+1}}}{\rd P_{X_i}}
              (\omega)\right) \rd\mu_{\pi}(\omega) \ge 0, 
\label{filtration-006}
\end{equation}
or equivalently,

\begin{equation}
   H\big[X^{ss}\|X_i\big] \ge  H\big[X^{ss}\|X_{i+1}\big] \ge 0, 
\label{filtration-007}
\end{equation}
where $X^{ss}:\mathscr{S}\to S$ is a random variable distributed according to the stationary distribution $\pi$.  This is essentially equivalent to the result in Eq. \ref{eq9}.  

Eq. \ref{filtration-006} states that information lost from
$X_i$ to $X_{i+1}$, averaged with respect to the invariant density, 
is always greater than zero, while Eq. \ref{filtration-007} suggests that
``the infinitely distant future has more information to gain from $X_i$ than 
from $X_{i+1}$''.  There is a subtle difference between 
these statements and the following: ``when including the future, the
world is at least as random as up to now.''  The reason for this, we 
suggest, is that (\ref{filtration-006}) and (\ref{filtration-007}) 
require the existence of the stationary measure.  Knowing 
the existence of a stationary behavior, ``the future is at least as 
random as now.''

\section{Deterministic correspondence and infinite $\beta$}
\label{sec:3}
Any representation of reality requires elements of  both chance and determinism.  These
correspond to the stochastic and deterministic 
components of complex dynamics.  As repeatedly pointed out
in \cite{haken-book-1,haken-book-2,haken-book-3}, it is the 
interaction between these two that yields self-orgranization
and complex behavior.  Therefore, the ability to ``envision''
a corresponding deterministic dynamics to some given stochastic dynamics, 
even when there is no obvious ``system size parameter'',
provides a deeper understanding of complex dynamics.
The natural parameter for a stochastic differential
equation (SDE) $\rd\x(t) = b(\x)\rd t + a\rd B(t)$ is the
noise strength $a$; the natural parameter in classical
statistical mechanics is the system's size (or one could use the temperature); and the natural
parameter in a Delbr\"{u}ck-Gillespie process is the
system's volume.

	How can one envision such a deterministic correspondence when no obvious natural parameters exist?  
It is becoming increasingly common to use the modal value of a 
distribution as a ``deterministic'' counter part to the stochastic system.  According to 
this view, a bimodal distribution corresponds to a bistable
system.   Note it is a widely held misconception that the 
mean dynamics $\langle \x(t)\rangle$
are the deterministic counterpart of a stochastic $\x(t)$.
For a SDE, $\langle \rd\x(t)\rangle \neq b(\langle\x\rangle)$
in general.  More importantly, while $\langle\x(t)\rangle$ is a non-random 
function of $t$, it is {\em not} a trajectory of any meaningful,
self-contained dynamical
system.  This point is best illustrated by the fact that the
differential equation describing $\langle\x(t)\rangle$ usually
depends on higher moments like 
$\langle\x^2(t)\rangle$.  Moreover, for a discrete system, even if 
the mean is defined, it does not usually lie in
the same space as $\x(t)$. 

We propose the following ``deterministic'' counterpart for a random variable 
$\x$ with probability mass function $p_{\x}^{ss}$, and we will show that it is intimately related to the energy defined in \eqref{eq:H}.  We will define the ``deterministic'' variable $\x_{\infty}$ as 

\begin{equation} \label{eq:determ}
\x_{\infty} = \lim_{\beta\to\infty}\x_{\beta}, 
\end{equation}
where 

\begin{equation}
\label{eq:betafam}
p^{ss}_{\x_{\beta}}(x) = \frac{p^{ss}_{\x}(x)^{\beta}}{Z(\beta)}, 
\end{equation}
with normalization constant

\[
Z(\beta) = \sum_{x}p^{ss}_{\x}(x)^{\beta}.  
\]
The random variable $\x_{\infty}$ will be concentrated on a finite number of states (the most probable
ones of $p_{\x}(x)$) with probability 1.  In particular, if $p^{ss}_{\x}(x)$ is unimodal, then $\x_{\infty}$ really will be a deterministic system.  On the other hand, if $p^{ss}_{\x}(x)$ is multimodal, then there is no unique deterministic counterpart.  Applying this idea to a 
discrete-state Markov process, the corresponding dynamics 
become a deterministic transformation, as discussed in \cite{ywq-dcds-b}.  

It is worth noting that similar definitions are often introduced formally as analogues to inverse-temperature without any discussion of deterministic correspondence (e.g., \cite{bialek-1, bialek-2}).  We spend so much time on the concept in order to emphasize that it arises naturally in a study of stochastic systems, without any reference to thermodynamic concepts.  The scaling factor $\beta$ should not just be thought of as a formal method for introducing temperature to a system, but as a natural feature of any probabilistic system.  

With this definition in hand, the obvious question becomes ``how fast does the limit in \eqref{eq:determ} converge?''  In the next section, we will try to make this question more rigorous, and in the process provide more evidence that $H(x)$ is an important quantity.  

\subsection{Large deviation principle for infinite $\beta$} \label{sec31}
We will now investigate the rate of convergence of the limit in \eqref{eq:determ}.  This is a question well suited to the methods of large deviation theory, but before we can use such methods we need to frame the question somewhat more rigorously.  Strictly speaking, we should be dealing with limits of measures rather than limits of random variables.  

Let $\left(\mathscr{S},\mathcal{F},P\right)$ be a discrete probability space with probability mass function $p^{ss}$ and define the family of measures $P^{\beta}$ on $\left(\mathscr{S},\mathcal{F}\right)$ whose probability mass functions are given by 

\begin{eqnarray}
p(x,\beta) &=& \frac{p^{ss}(x)^{\beta}}{Z(\beta)}, \textrm{ where } \nonumber\\
Z(\beta) &=& \sum_{x\in\mathscr{S}}p^{ss}(x)^{\beta}.  
\end{eqnarray}
(As we will show later, this is always possible for $\beta \geq 1$.)  In addition, let $\left(S,\Sigma\right)$ be a measurable space and choose a function $\sigma:\mathscr{S}\to S$.  This defines a family of $S$-valued random variables $\mathcal{O}_{\beta}$, where 

\begin{equation}
\textrm{Pr}\left\{ \mathcal{O}_{\beta} = z \right\} = P^{\beta}\left(\left\{ x\in\mathscr{S} \:\vert\: \sigma(x) = z\right\}\right).  
\end{equation}
In particular, if $\sigma$ is one-to-one, then $\textrm{Pr}\left\{\mathcal{O}_{\beta} = z\right\} = p(\sigma^{-1}(z),\beta)$.  

For unimodal distributions, we know that as $\beta$ goes to infinity, the distribution of $\mathcal{O}_{\beta}$ becomes concentrated on a single value $z^{*}\in S$.  However, it is not clear \emph{a priori} how the rate of this convergence depends on our choice of $\mathcal{O}$.  It is conceivable that different observables could lead to different convergence rates.  Moreover, we could eschew observables altogether and work solely with the measures $P^{\beta}$.  In this section we will show that the rate of convergence is identical for a wide range of observables, and that it is intimately related to $H(x)$.  

\textbf{Case (i)}: Let $S = \R$.  We will not restrict $\sigma$ to be one-to-one, but we will assume that if $\sigma(x_1) = \sigma(x_2)$ for some $x_1,x_2\in\mathscr{S}$, then $p^{ss}(x_1) = p^{ss}(x_2)$.  We will let $N(x)$ denote the (necessarily finite) number of elements $y\in\mathscr{S}$ such $\sigma(y) = \sigma(x)$.  Finally, let $x^{*}\in\mathscr{S}$ be a state with maximal probability.  We know that 

\begin{equation}
\lim_{\beta\to\infty}\textrm{Pr}\left\{ \absval{\mathcal{O}_{\beta} - \sigma(x^{*})} \geq \eta \right\} = 0
\end{equation}
for any $\eta\in\R^{+}$.  In fact, $\textrm{Pr}\left\{ \absval{\mathcal{O}_{\beta} - \sigma(x^{*})} \geq \eta \right\}$ is a non-increasing step function of $\eta$.  Under reasonable conditions, we can write 

\begin{equation}
\textrm{Pr}\left\{ \absval{\mathcal{O}_{\beta} - \sigma(x^{*})} \geq \eta \right\} = e^{-\beta I_1(\eta) + o(\beta)}, 
\end{equation}
where 

\begin{equation}
I_1(\eta) = -\lim_{\beta\to\infty}\ln \textrm{Pr}\left\{ \absval{\mathcal{O}_{\beta} - \sigma(x^{*})} \right\}.  
\end{equation}
If we define $\hat{x}_{\eta} = \argmax_{x\in\mathscr{S}}\left\{ \absval{\sigma(x) - \sigma(x^{*})} \right\}$, then we have 

\begin{align*}
I_1(\eta) &= -\lim_{\beta\to\infty}\frac{1}{\beta}\ln\left(\frac{1}{Z(\beta)}\sum_{x:\absval{\sigma(x) - \sigma(x^{*})}\geq\eta}p^{ss}(x)^{\beta}\right) \\
&= -\lim_{\beta\to\infty}\frac{1}{\beta}\ln\left(\frac{N(\hat{x}_{\eta})p(\hat{x}_{\eta},\beta)}{N(x^{*})p(x^{*},\beta)}\right) \\
&= -\ln\left(\frac{p(\hat{x}_{\eta},1)}{p(x^{*},1)}\right) \\
&= H(\hat{x}_{\eta}) - H(x^{*}).  
\end{align*}

\textbf{Case (ii)}: Instead of creating a somewhat arbitrary family of observables $\mathcal{O}_{\beta}$, we can also work solely with the measures $P^{\beta}$.  To make this more convenient, we will introduce some additional notation.  

Let $\mathscr{Y} = H(\mathscr{S})\subset\R$ and let $y^{*}$ be the minimum value in $\mathscr{Y}$.  For any $h > y^{*}$, let $\mathscr{S}_{h} = \left\{ x\in\mathscr{S} \:\vert\: H(x) < h \right\}$ and $\mathscr{Y}_{h} = \left\{ y\in\mathscr{Y} \:\vert\: y < h \right\}$.  Let $\left\lfloor h \right\rfloor$ denote the minimum value of $\mathscr{Y}\setminus\mathscr{Y}_h$.  Finally, define 

\begin{equation}
Z_h(\beta) = \sum_{x\in\mathscr{S}_h}p^{ss}(x)^{\beta}.  
\end{equation}
We know that $P^{\beta}\left(\mathscr{S}\setminus\mathscr{S}_h\right)$ approaches zero as $\beta$ goes to infinity.  Much like the previous case, we would like to know how quickly this quantity decays.  We have 

\begin{equation}
P^{\beta}\left(\mathscr{S}\setminus\mathscr{S}_h\right) = e^{-\beta I_2(h) + o(\beta)}, 
\end{equation}
where 
\begin{align*}
I_2(h) &= -\lim_{\beta\to\infty}\frac{1}{\beta}\ln P^{\beta}\left(\mathscr{S}\setminus\mathscr{S}_h\right) \\
&= -\lim_{\beta\to\infty}\frac{1}{\beta}\ln\left(\frac{1 - Z_h(\beta)}{Z(\beta)}\right) \\
&= -\lim_{\beta\to\infty}\frac{1}{\beta}\ln\left(\frac{\sum_{x:-\ln p^{ss}(x) = \lfloor h \rfloor}p^{ss}(x)^{\beta}}{\sum_{x:-\ln p^{ss}(x) = y^{*}}p^{ss}(x)^{\beta}}\right) \\
&= \lfloor h \rfloor - y^{*}.  
\end{align*}

In fact, this is in some sense just a special case of case (i).  If we choose $\sigma = H$ and let $h = \eta + y^{*}$, then $I_1$ and $I_2$ are identical.   

{\bf Case (iii)}: One of the key insights from the theory of large deviations
is that in the limit of $\beta\rightarrow\infty$, the
probability $\Pr\big\{ \x_{\beta} \notin \mathscr{S}_h\big\}$
is determined by one particular $x^{*}\notin\mathscr{S}_h$,
the one with $p(x^{*},1)\ge p(x,1)$ $\forall x\notin\mathscr{S}_h$.
Therefore, one has $\lim_{\beta\to\infty}p(x,\beta)\approx e^{-\beta I_3(x)}$, for any $z\in\mathscr{S}$.  This is essentially the same as the WKB ansatz.  We then have 

\begin{eqnarray}
    I_3(x)  &=& -\lim_{\beta\rightarrow\infty}
              \frac{1}{\beta}\ln p(x,\beta)
\nonumber\\
  &=&   -\ln p(x,1)
          + \lim_{\beta\to\infty} \frac{1}{\beta} 
            \ln\sum_{x\in\mathscr{S}} p^{\beta}(x,1)
\nonumber\\
	&=&  -\ln\left(\frac{p(x,1)}{p(x^*,1)}\right)
          + \lim_{\beta\to\infty} \frac{1}{\beta} 
            \ln\left[ 1 + \sum_{x\in\mathscr{S},x\neq x^*} 
         \left(\frac{p(x,1)}{p(x^*,1)}\right)^{\beta}
            \right]
\nonumber\\
	&=&  -\ln\left(\frac{p(x,1)}{p(x^*,1)}\right)
\nonumber\\
&=& H(x) - H(x^{*}). 
\nonumber 
\end{eqnarray}

\section{Entropy, energy and criticality in systems with generalized potential}
\label{sec:criticality}

	The results of the previous section suggest that $H(x) = -\ln p^{ss}(x)$ is a mathematically relevant quantity and that it can reasonably be interpreted as an energy.  We will now investigate some of the consequences of this definition in more detail.  In particular, we will shed some light on the distinction between Gibbs and Boltzmann entropies and derive a necessary and sufficient condition for the existence of a critical temperature in stationary stochastic systems.  

Let us again suppose that our system takes on possible
states from a discrete (finite or countably infinite) set $\mathscr{S}$, and let $p^{ss}:\mathscr{S}\to[0,1]$ be the probability mass function describing the chance that event $x\in\mathscr{S}$ ocurrs.  As above, we will define the energy of a state $x\in\mathscr{S}$ as 

\begin{equation} \label{eq:H-sec3}
H(x) = -\ln p^{ss}(x), 
\end{equation}
In addition, we will avoid substantial difficulties later if we endow $H$ with units of energy.  If we do so, then we can no longer simply write $p^{ss}(x) = e^{-H(x)}$.  Instead, we need to introduce another parameter $\beta$ with units of inverse energy.  This gives us

\begin{equation} \label{eq:boltzmann}
p^{ss}(x;\beta) = \frac{1}{Z(\beta)}e^{-\beta H(x)},
\end{equation}
where the partition function $Z(\beta)$ is defined as 

\begin{equation} \label{eq:part}
Z(\beta) = \sum_{x\in \mathscr{S}}e^{-\beta H(x)}
\end{equation}
Note that the partition function is necessarily a dimensionless quantity, as discussed in \cite{sack, munster, brown}.  These distributions are precisely the probability mass functions of the measures $P^{\beta}$ defined in Sec. \ref{sec:3}.  

With this definition, there is a serious concern that the sum in \eqref{eq:part} might not converge.  Since $p^{ss}$ is a probability distribution, however, we do know that the sum converges for $\beta = 1$ (in fact, we know that $Z(1) = 1$.)  We will spend much of the following sections discussing the cases where the sum in \eqref{eq:part} diverges, but for the moment we will simply assume that $Z(\beta)$ is well-defined on some subset of $\R$ containing $[1,\infty)$. 

In classical statistical mechanics, one typically has the mechanical 
energy function in hand before $p^{ss}$, and then shows that the 
system at finite ``temperature'' $\beta^{-1}$ has an equilibrium 
distribution among the states described by \eqref{eq:boltzmann}. 
Note that when $\beta\rightarrow\infty$, the distribution 
$p^{ss}(x;\beta)$ converges to a uniform probability distribution on the set of states with minimal $H$.  For certain non-convex $H(x)$, the 
phenomenon of phase transition occurs \cite{aqtw}.  This limit gives precisely the deterministic correspondence described in Sec. \ref{sec:3}.  

In a classical statistical mechanical problem, $\mathscr{S}$ is a continuous space describing the positions and momenta of all particles in the system, $H$ is a Hamiltonian for this system and $\beta = (k_{B}T)^{-1}$ is the inverse temperature.  One would then be interested in level sets with constant energy $h$.  In particular, Gibbs' and Boltzmann's entropies are concerned with the phase volume and phase surface area of such level sets.  

Unlike in a classical problem, though, our state space $\mathscr{S}$ is arbitrary, and in general may not be useful as a phase space.  In particular, $\mathscr{S}$ often does not come equipped with a metric, or even any sort of order.  To remedy this, we will define the rank of a state $x$ as 

\begin{equation}
\mathfrak{R}(x) = \#\big|\big\{y\in \mathscr{S} \:\vert\: H(x) \geq H(y)\big\}\big|, 
\end{equation}
where $\#\absval{\cdot}$ denotes cardinality.  That is, the rank of $x$ is the number of states which have lower energy than $x$ (or are at least as probable as $x$).  Since $\mathfrak{R}$ depends on $x$ only through $p^{ss}(x)$, we can unambiguously define the rank in terms of energy as $\mathcal{V}:[0,\infty)\to\Z^{+}$ as 

\begin{equation}
\mathcal{V}(h) = \#\big|\big\{x\in \mathscr{S} \:\vert\: H(x) \leq h\big\}\big|, 
\end{equation}
so that $\mathfrak{R}(x) = \mathcal{V}(H(x))$ for every $x\in \mathscr{S}$.  

Notice that $\mathcal{V}$, as opposed to $\mathfrak{R}$, is no longer defined on a discrete space -- it is a function of the continuous variable $h$ -- but because $\mathscr{S}$ is discrete, $\mathcal{V}$ can be written as a non-decreasing piecewise constant function.  

It is also worth noting that our assumption of a countable state space cannot be easily relaxed in this approach.  If $\mathscr{S}$ were uncountable, then one could not hope to order the states by their rank.  Indeed, $\mathfrak{R}$ and $\mathcal{V}$ would generally be infinite for almost all input.  Such issues arise because $p^{ss}$ is, by assumption, a probability density with respect to the counting measure.  We could have instead assumed that $p^{ss}$ was a density with respect to some other measure (e.g., the Lebesgue measure on $\mathscr{S} = \R$), but this would introduce many other subtleties later on.  

\subsection{Microcanonical partition functions and entropy}
\label{sec41}

If we take the liberty of treating the derivative of a Heaviside 
function as a Dirac$-\delta$ function, then we can write $\mathcal{V}$ as 

\begin{equation}
\mathcal{V}(h) = \int_{0}^{h}\d\mathcal{V}(y) = \int_{0}^{h}\pderiv{\mathcal{V}}{y}\d y.  
\end{equation}
It is very important to note that $\partial\mathcal{V}(h)/\partial h$ has units of inverse energy.  It is tempting (and often quite useful) to define 

\begin{equation}
\Omega(h) = \#\big|\big\{x\in \mathscr{S} \:\vert\: H(x) = h\big\}\big|,
\end{equation}
and then write 

\begin{equation}
\mathcal{V}(h) = \sum_{n=0}^{\infty}\Omega(h_n), 
\end{equation}
where the sum is taken over the values $h_n \leq h$ such that $\Omega(h_n) > 0$.\footnote{For any finite $h$, note that $\mathcal{V}(h) \leq e^h$ because the distribution $p^{ss}$ sums to 1.  The number of distinct values $h_n \leq h$ is no greater than $\mathcal{V}(h)$, so it too is finite.}  However, one should keep in mind that $\d\mathcal{V}/\d h \neq \Omega(h)$.  That is, $\d\mathcal{V}/\d h$ is not really just a number of states; it is a density\footnote{This is a common issue in probability theory as well.  The probability of an event $A$ should always be written as $\int_{A}\d F = \int_{A}(\rd F/\rd x)\d x = \int_{A}f(x)\d x$, where $F$ is the cumulative probability measure and $f =\rd F/\rd x$ is a density with respect to some other measure.  When the other measure is a counting measure, however, it is commonplace to replace the integral with a sum and use the probability mass $p(x) = (\rd F/\rd x)\d x$ instead.  This is \emph{numerically} correct, but often leads to confusion over units.}.

One of the main reasons we have introduced this notation with $\mathcal{V}$ is that it gives us a much more convenient way to write $Z(\beta)$.  In particular, we can write $Z$ without reference to the individual states $x$.  

\begin{equation}
Z(\beta) = \sum_{x\in \mathscr{S}}e^{-\beta H(x)} = \int_{0}^{\infty}e^{-\beta h}\d\mathcal{V}(h), 
\end{equation}
This is exactly the Laplace-Stieltjes transform of $\mathcal{V}$.  

It is tempting to rewrite $Z$ as 

\begin{equation} 
\label{eq:boltzmann-part}
Z(\beta) = \int_{0}^{\infty}e^{-\beta h}\left(\pderiv{\mathcal{V}}{h}\right)\d h = \int_{0}^{\infty}e^{-\beta\left(h - (k_{B}\beta)^{-1}k_{B}\ln\Omega(h)\right)}\d h, 
\end{equation}
and to then identify $\partial\mathcal{V}/\partial{h}$ as the microcanonical partition function and $k_{B}\ln\Omega(h)$ as the entropy.  Unfortunately, this is entirely wrong.  Equation \ref{eq:boltzmann-part} relies on the identification of $\pderiv{\mathcal{V}}{h}$ with $\Omega(h)$, which is invalid.  This method can be salvaged by introducing a factor $\Delta h$ with units of energy, so that the \eqref{eq:boltzmann-part} becomes 

\begin{equation}
Z(\beta) = \int_{0}^{\infty}\frac{1}{\Delta h}e^{-\beta h}\left(\Delta h\pderiv{\mathcal{V}}{h}\right)\d h, 
\end{equation}
and the entropy becomes 

\begin{equation} \label{eq:boltzmann-ent}
S_{B} = k_{B}\ln\left(\Delta h\pderiv{\mathcal{V}}{h}\right).  
\end{equation}
In fact, if we choose $\Delta h$ as a constant, then this is exactly the Boltzmann entropy.  Such a solution is somewhat unsatisfying; the introduction of arbitrary constants to correct units generally suggests a deeper misunderstanding.  (Worse yet, there is no real reason for $\Delta h$ to be constant, so long as it has the correct units.)  

A much more satisfying interpretation of $Z$ arises if we integrate by parts, obtaining 

\begin{equation} \label{eq:gibbs-part}
Z(\beta) = \beta\int_{0}^{\infty}e^{-\beta h}\mathcal{V}(h)\d h = \beta\int_{0}^{\infty}e^{-\beta\left(h - (k_{B}\beta)^{-1}k_{B}\ln\mathcal{V}(h)\right)}\d h.  
\end{equation}
Here, we can interpret $\mathcal{V}(h)$ as the microcanonical partition function and

\begin{equation} \label{eq:gibbs-ent}
S_{G} = k_{B}\ln\mathcal{V}(h)
\end{equation}
as the entropy.  We have chosen the subscripts $G$ and $B$ to emphasize that $S_{B}$ corresponds to Boltzmann entropy, while $S_{G}$ corresponds to Gibbs entropy.  

There has been much debate over the relative merit of these definitions of entropy in statistical mechanics (e.g., \cite{campisi, jaynes, dunkel-hilbert, frenkel-warren}).  While we do not claim to have resolved this question, equations \eqref{eq:boltzmann-part} and \eqref{eq:gibbs-part} suggest that Gibbs entropy is the more natural choice.  Furthermore, as we will see in the next section, Gibbs entropy plays a central role in the notion of criticality.  

It is worth noting that the terminology surrounding Boltzmann and Gibbs entropy is not entirely consistent.  Most notably, some authors (e.g.,  \cite{goldstein-lebowitz, lebowitz}) use the phrase ``Boltzmann entropy'' to refer to the logarithm of the volume of any phase space region corresponding to a suitable macrostate and use ``Gibbs entropy'' to refer to the quantity $\int p\ln p\d x$, where $p$ is some probability density.  Using this terminology, \eqref{eq:boltzmann-ent} and \eqref{eq:gibbs-ent} would both be Boltzmann entropies, but would use different macrostates.  

Instead, we follow the convention used in, e.g., \cite{jaynes, campisi, dunkel-hilbert, frenkel-warren} and use ``Boltzmann entropy'' to indicate the logarithm of the volume of a thin shell in phase space and ``Gibbs entropy'' to indicate the logarithm of the volume of the interior of such a shell.  If the quantity $\int p\ln p\d x$ is needed, we will refer to it as Shannon entropy.

\subsection{Analyticity of $Z$ as a function of $\beta$}
\label{crit-sec}

	The analyticity of $Z(\beta)$, which is analogous to the partition function
in statistical mechanics, is intimately related to phase transitions and
critical phenomena \cite{zimm,yang-lee,lee-yang,blythe-evans}.
Our system has a critical temperature (in the statistical mechanical sense of the term) if and only if the partition function is non-analytic for some $\beta\in(0,\infty)$.  Since $Z(\beta)$ is a Laplace transform, we have access to some useful theorems from classical analysis, all of which can be found in \cite{widder}.  

First, there is some value $\beta_c\in[-\infty,\infty]$ such that $Z(\beta)$ converges for all $\beta\in\mathbb{C}$ with real part greater than $\beta_c$ and diverges for all $\beta\in\mathbb{C}$ with real part less than $\beta_c$.  The value $\beta_c$ is called the abscissa of convergence.  

Second, if the state space $\mathscr{S}$ is finite then $Z$ is a sum of finitely many terms and therefore converges for any $\beta$ (i.e., $\beta_c = -\infty$).  However, if $\mathscr{S}$ is infinite then the partition function will not be analytic for all real $\beta$.  In particular, it cannot converge when $\beta = 0$ because $Z(0) = \#\absval{\mathscr{S}}$.  However, by definition we know that $Z(\beta)$ converges when $\beta = 1$, since $Z(1)$ is the normalization constant of $p^{ss}$.  For infinite systems, the abscissa of convergence must therefore lie somewhere in $[0,1]$.  

Since the abscissa of convergence is non-negative, we have

\begin{equation}
\beta_c = \limsup_{h\to\infty}\frac{\ln\mathcal{V}(h)}{h}, 
\end{equation}
or 

\begin{equation} \label{eq:betac}
k_{B}\beta_c = \limsup_{h\to\infty}\frac{S_G(h)}{h}.  
\end{equation}
We now know that the partition function is analytic for all complex $\beta$ with real part greater than $\beta_c$, where $\beta_c$ is found as in \eqref{eq:betac}.  However, we have not yet shown that $Z(\beta)$ cannot be extended analytically beyond $\beta = \beta_c$.  For a general Laplace-Stieltjes transform, this might be possible.  (In the worst case, a Laplace transform may have a finite abscissa of convergence, but still have an analytic continuation to the entire complex plane.)  Fortunately, since $\mathcal{V}$ is monotonic, $Z(\beta)$ has a singularity at $\beta_c$.  (This also means that $\beta_c \neq 1$.)  

This means that the partition function $Z(\beta)$ has a singularity at some positive $\beta_c$ if and only if $S_G$ is asymptotic to $h$ in the sense of \eqref{eq:betac}.

\subsection{Examples}
\label{sec43}

So far, we have let our system be very general.  The arguments above apply equally well to a wide range of systems -- from the single electron of a hydrogen atom (where $\mathscr{S}$ is the set of possible orbits) to the configuration of amino acids in a strand of DNA.  It is not immediately clear how \eqref{eq:betac} might be influenced by the structures of $\mathscr{S}$ and $p^{ss}$.  To illustrate the consequences of our result, we will look at a few examples.  

First, we will investigate two so-called ``non-degenerate'' cases where each state has a distinct probability (i.e., $\Omega \equiv 1$).  Since we only care about the rank of states, we will suffer no loss of generality by assuming that $\mathscr{S} = \Z^{+}$ and that the states are ordered so that $p^{ss}(x) > p^{ss}(y)$ whenever $x < y$.  As an example, consider the distribution: 

\begin{equation}
p^{ss}(x) = 2^{-x}.  
\end{equation}
We have 

\begin{eqnarray}
H(x) &=& x\ln 2, 
\nonumber\\
S_{G}(h) &=& k_B\ln h - k_B\ln\ln 2, \:\textrm{ and } 
\nonumber\\
\beta_c &=& 0.   
\end{eqnarray}
This distribution therefore does not have a critical temperature, which should not be surprising, since it is exponential.  

Alternatively, consider a power law distribution.  

\begin{equation}
p^{ss}(x) = \frac{x^{-\alpha}}{\zeta(\alpha)}, 
\end{equation}
where $\alpha > 1$ and $\zeta$ is the Riemann zeta function.  This gives us 

\begin{eqnarray}
H(x) &= \alpha\ln x + \ln\zeta(\alpha), 
\nonumber\\
S_{G}(h) &= \frac{k_B}{\alpha}\left(h -  \ln\zeta(\alpha)\right) \:\textrm{ and } 
\nonumber\\
\beta_c &= \frac{1}{\alpha}.  
\end{eqnarray}
This means that power law distributions do indeed have a critical temperature.  This result was already demonstrated in \cite{bialek-1}, but arises as a special case of our work.  

These examples highlight the main feature of criticality: a system will be critical if and only if the probability of a state decays too slowly as a function of rank.  That is, critical distributions are fat-tailed in ``phase space''.  

We observe a similar result when $\Omega$ is not identically 1 (``degenerate'' distributions).  For example, consider a distribution where, for each $n\in\Z^{+}$, there are $2^{n}$ states with stationary probability $2^{-2n}$.  That is, for each $h_n = 2n\ln 2$, we have $\Omega(h_n) = 2^{n}$.  In this case, 

\begin{equation}
\mathcal{V}(h) = 2\left(2^{n} - 1\right) \mbox{ for } h_n\leq h < h_{n+1}, 
\end{equation}
and we find that $\beta_c = 1/2$.  In light of our previous examples, this should not be surprising: when written as a function of rank, $p^{ss}$ decays like $x^{-2}$, so this $\beta_c$ is exactly what we expect.  However, it also illustrates the importance of how we label our state space.  

Suppose that we observed the system given above, but that we could not identify each individual state.  If instead of observing $2^{n}$ distinct states, each with probability $2^{-2n}$, we only measured 1 state with probability $2^{-n}$, we would then calculate the probability distribution $p^{ss}(x) = 2^{-x}$, for which $\beta_c = 0$.  Depending on how states are counted, the distribution could either have a critical temperature or not!  This distinction is exactly why the partition functions in classical and quantum statistical mechanics differ by a factor of $N!$.  The classical version overcounts the number of possible microstates because it assumes particles are distinguishable.  Without the correction term, this would often lead to substantially different predictions between the two theories.  Fortunately, we know that quantum mechanics is the correct theory, and so we are able to choose the correct definition of a microstate.  

In many applications, however, we do not know what a true microstate looks like.  For example, imagine a particle undergoing a random walk on a lattice $X$, and suppose that we can measure only the distance $r$ between a particle and the origin.  It would be natural to define a microstate of this system by the distance between the particle and the origin.  If $X = \Z^{+}$, then this is exactly correct, but if $X = \Z$, then there are really 2 microstates for each $r$.  Worse yet, if the lattice is two-dimensional (i.e., $X = \Z\times\Z$), then each $r$ corresponds to a different number of microstates, and this number grows without bound.  As discussed in section \ref{sec:2.1}, we can still find a reasonable interpretation for the energy of such a system.  If we treat each $r$ as a microstate, then $H(r)$ is the potential of mean force in the radial direction.  However, our notions of entropy and criticality may change drastically depending on how we define our state space.  

For a slightly more involved example, consider the so-called ``zipper model'' (described in, e.g., \cite{gibbs-dimarzio, kittel, nagle}).  This is a highly simplified model of, among other things, the conformation of a double-stranded DNA molecule.  Suppose there are $N$ base pairs along the DNA molecule (where $N$ can be a positive integer or $\infty$; if $N = \infty$ then think of the molecule as having a fixed left end, but extending infinitely to the right), each of which can either be linked or broken.  We will assume that there is only one possible linked configuration for each base pair, but that there are $G$ possible broken configurations for each pair, where $G$ is a positive integer.  Furthermore, we will suppose that bonds are only broken from left to right.  That is, it is possible for a base pair to be in one of the $G$ broken configurations if and only if every base pair to the left is also broken.\footnote{Allowing the bonds to break from both ends makes the formulas that follow somewhat more complicated, but does not qualitatively alter the behavior of the system.  On the other hand, allowing arbitrary bonds to be broken will make the state space of our system uncountable when the chain becomes infinite.  As we will discuss in the next section, this has important consequences.}  Suppose that the energy of a linked base pair is 0 and that the energy of any of the $G$ broken configurations for a single base pair is $E > 0$ if all base pairs to the left are broken and infinite otherwise.  When $N = \infty$ and $G > 1$, this system has a phase transition at $\beta = \ln G/E$.  Otherwise, it has no critical temperature \cite{kittel}.  We will show that this critical behaivor is reproduced using \eqref{eq:betac}.  

The state space $\mathscr{S}$ of this system is the collection of all possible allowed configurations of linked and broken base pairs.  Each configuration consists of $m$ broken base pairs followed by $N - m$ linked base pairs, and there are $G^{m}$ distinct states for each $m$.  Notice that $\mathscr{S}$ is finite whenever $N$ is and countably infinite when $N = \infty$.  The probability of each of these configurations is given by 

\begin{equation}
p^{ss}(x; N) = \frac{1}{Q_N}e^{-m E}, 
\end{equation}
where $m$ is the number of broken base pairs in $x$ and $Q_N$ is a constant that depends on $N$ (but not $x$).  Note that it is not immediately obvious from the previous assumptions that $p^{ss}(x; N)$ is well-defined, but one can show that $Q_{\infty}$ is non-zero and finite for sufficiently large $E$.  (In fact, we can solve for $Q_{\infty}$ exactly, but for our purposes it is enough to know that it is finite.)  

Since $\mathscr{S}$ is finite whenever $N$ is, we know that there is no critical temperature for $p^{ss}(\cdot; N)$ when $N < \infty$, so consider the case where $N = \infty$.  The possible energy values are $h_m = mE - \ln Q_{\infty}$ for any $m\in\Z^{+}$.  The Gibbs entropy is therefore 

\begin{equation}
S_G(h_m) = k_B\ln\left(\sum_{k=0}^{m}G^k\right) = k_B\ln\left(\frac{G^{m+1} - 1}{G - 1}\right), 
\end{equation}
if $G > 1$ and $S_G(h_m) = k_B\ln\left(m+1\right)$ if $G = 1$.  

Applying \eqref{eq:betac}, we therefore have 

\begin{equation}
\beta_c = \limsup_{m\to\infty}\frac{\ln\left(G^{m+1} - 1\right) - \ln\left(G - 1\right)}{mE - \ln Q_{\infty}} = \lim_{m\to\infty}\frac{(m+1)\ln G}{mE} = \frac{\ln G}{E}, 
\end{equation}
when $G > 1$.  If $G = 1$, we have 

\begin{equation}
\beta_c = \limsup_{m\to\infty}\frac{\ln\left(m+1\right)}{mE - \ln Q_{\infty}} = 0.  
\end{equation}

These critical temperatures exactly match the known values, and the mechanism for this behavior is easy to see.  When $G = 1$, the phase-volume $\mathcal{V}(h)$ grows linearly with $h$, but when $G > 1$ the phase-volume grows exponentially.  This allows the entropy $S_G$ to keep pace with the energy as $h$ grows, leading to a criticality.  

The preceding calculations are quite similar to those used in the equilibrium statistical mechanical approach of Kittel (\cite{kittel}), but the procedure is very different in spirit.  In Kittel's approach, one finds $Q_N$ for arbitrary $N$, then uses $Q_N$ to calculate a statistic such as the expected number of broken base pairs.  Finally, one takes the limit as $N\to\infty$ and demonstrates that this statistic becomes non-analytic at some finite temperature.  In particular, Kittel warns that ``it is dangerous to write ... the partition function for $N = \infty$; the correct procedure is to evaluate the thermodynamic quantities for finite $N$ and then to examine the limit.''  In our approach, we start by finding $p^{ss}(x; \infty)$ (up to a constant).  Once we have obtained this distribution, we can calculate $S_G(h)$ for the infinite system and directly obtain $\beta_c$.  The danger that Kittel describes is still present: our method will fail if $p^{ss}(x; \infty)$ is not well-defined.

\section{Discussion}
\label{sec:5}

It is worth taking a moment to discuss not only what we have shown in the previous sections, but also what we have not shown.  We have demonstrated that a stationary distribution over a discrete state space has a critical temperature if and only if the Gibbs entropy of the distribution \eqref{eq:gibbs-ent} satisfies the relation \eqref{eq:betac}.  The terminology used here is deliberately suggestive, but one should not take it too far.  For one thing, there are phase transitions in equilibrium statistical mechanics that do not seem to fit the description given in section \ref{sec:criticality}.  The Lee-Yang theorem, for instance, describes cases where the partition function becomes zero rather than infinite, and two-dimensional Ising models can exhibit different types of phase transitions.  

The key point is that we have assumed, from the outset, the existence of a well-defined stationary probability distribution on a countable state space.  Such a distribution has a critical temperature $\beta_c$ if $Z(\beta)$ approaches either zero or infinity as $\beta \to \beta_c$.  Because $p^{ss}(x;\beta=1)$ is a probability mass function, $Z(\beta)$ cannot become zero for any finite $\beta$.  That is, Lee-Yang type criticalities can only occur if the stationary distribution $p^{ss}$ is not well-defined for any temperature.  

Ising models, on the other hand, may have well-defined equilibrium distributions even in the thermodynamic limit.  However, these models typically have an uncountable state space when $N\to\infty$.  For such a distribution, the proofs of section \ref{sec:criticality} do not hold as written and other types of criticalities may be present.  

Mora and Bialek have also discussed this approach in regards to Ising models \cite{bialek-1}.  In particular, they showed that systems where $p^{ss}(x; N) \propto \mathfrak{R}(x)^{-\alpha}$ follows a power law have a critical temperature given by $\beta_c = 1/\alpha$ when $N$ goes to infinity.  Their result utilized the identification of $S_G$ with $S_B$, which becomes precise in the thermodynamic limit.  In the present paper, we have shown that such an identification is unnecessary and that the critical temperature conditions are still exact in ``smaller'' systems.  Moreover, we have found a broader condition for the existence of a critical temperature, of which the power law relationship is a special case.  

After Mora and Bialek's paper, there has been much discussion about the idea that biological systems are poised at a critical point.  This idea arose because researchers obtained estimates of $p^{ss}$ for a wide range of biological systems, and all appeared to follow some sort of power law.  Such a distribution would indicate a non-zero abscissa $\beta_c$.  The result from Sec. \ref{crit-sec} does seem like it should indicate a criticality, but there are some important caveats worth considering.  

First, it is notoriously difficult to calculate tail properties (such as $\beta_c$) from an estimated distribution.  Estimates of $p^{ss}$ are necessarily based on a finite number of samples, and therefore cannot give reliable information about arbitrarilly low probability events, which is required to calculate \eqref{eq:betac}.  

Second, and much more insidious, many biological processes are 
not in a true steady state.  The formal analogies we have made with statistical mechanics only make sense in the context of stationary systems.  If $p^{ss}$ is actually actually varies slowly with respect to some other variable (most importantly time), then our notion of criticality does not necessarily correspond to any interesting feature of the system.  For instance, Schwab, Nemenman and Mehta \cite{schwab} has shown that slowly varying latent variables can give rise to apparent power law distributions, which necessarily have a non-zero $\beta_c$, even in conditionally independent systems.




\end{document}